\documentclass[fleqn,usenatbib]{mnras}

\usepackage[T1]{fontenc}

\DeclareRobustCommand{\VAN}[3]{#2}
\let\VANthebibliography\thebibliography
\def\thebibliography{\DeclareRobustCommand{\VAN}[3]{##3}\VANthebibliography}

\usepackage{graphicx}	
\usepackage{amsmath}	
\usepackage{amssymb}	
\usepackage{bm}
\usepackage{marginnote}
\usepackage{natbib}

\title[parameters deviating from intrinsic value]{The effect of redshift degeneracy and the damping effect of viscous medium on the information extracted from gravitational wave signals}

\author[Shou-Li Ning,LiXin Xu]{Shou-Li Ning,$^{1}$
\thanks{Contact e-mail:1043903278@qq.com}
LiXin Xu,$^{2}$\thanks{Contact e-mail: lxxu@dlut.edu.cn}
\\
$^{1}$ $^{2}$Institute of Theoretical Physics, School of Physics, Dalian University of Technology,
 Dalian, 116024, China
}

\date{Accepted XXX. Received YYY; in original form ZZZ}

\pubyear{2015}

\begin{document}
\label{firstpage}
\pagerange{\pageref{firstpage}--\pageref{lastpage}}
\maketitle

\begin{abstract}
Considering the cosmological redshift $z_c$, the mass of GW source extracted from GW signal is $1+z_c$ times larger than its intrinsic value, and distance between detector and GW source should be regarded as luminosity distance. However, besides cosmological redshift, there are other kinds of redshifts should be considered, which is actually ignored, in the analysis of GW data, such as Doppler redshift and gravitational redshift, so the parameters extracted from GW may deviate from their intrinsic values. Another factor that may affect GW is the viscous medium in propagation path of GW, which may damp the GW with a damping rate of $16\pi G\eta$. Some studies indicate dark matter may interact with each other, thus dark matter may be the origin of viscosity of cosmic medium. Then the GW may be rapidly damped by the viscous medium that is made of dark matter, such as dark matter "mini-spike" around intermediate mass black hole. In this article, we mainly discuss how Doppler and gravitational redshift, together with the damping effect of viscous medium, affect the information , such as the mass and redshift of GW source, extracted from GW signals.
\end{abstract}
\begin{keywords}
gravitational waves -- dark matter.
\end{keywords}



\section{Introduction}

In general, one calculates the gravitational wave waveform in barycentric reference system. In this reference, the GW source is of course motionless. However, the detector and the source are usually not at rest relative to each other. For example, the speed of solar system rounding the center of the Milk Way is about $240km/s$, and the speed of solar system relative to CMB is $368\pm 2 km/s$. In view of cosmic expansion, the frequency of GW signals received by detector may have a redshift. And the parameters extracted from GW signals deviate from their intrinsic values, for example, the mass is $1+z_c$ times larger than its intrinsic value \cite{Krolak_1987}. The motion of GW source relative to detector gives rise to the shift of GW frequency too. So the motion of the GW source must be concerned if the speed of GW source is large enough, for instance, radiation-reaction of GW may speed the source up to $1000km/s-5000km/s$ \cite{Gonz_lez_2007,Lousto_2011}, and the corresponding Doppler redshift is $0.003-0.016$. If the wave source is in a deep gravitational potential well, then gravitational red shift and Doppler red shift, which are caused by GW source moving in potential well, are both important. Another factor that may affect the GW is the viscous medium in the propagation path of the wave. Because of the damping effect of medium, the intense of wave decreases when wave goes through the medium \cite{Baym_2017,Goswami_2017,Kocsis_2008}. The amplitude of GW is inversely proportional to the distance, so the distance extracted from GW is lager than its true value if the GW is damped. The damping effect, together with redshift, affects the waveform. These unfavourable factors may cause the parameters of GW source deviate from their intrinsic values.

In section~\ref{sec:2}, we mainly discuss the Doppler and gravitational redshift and how these two kinds of redshift affect the parameter estimation of GW. In section~\ref{sec:3}, we discuss how GW is damped by high-density dark matter and how the damping effect changes the parameters of GW source. 

\section{The effect of Doppler and gravitational redshift on GW}
\label{sec:2}

Here we assume that the velocity of the wave source relative to detector is $\bm{v}$, and $\bm{n}$ is unit vector aligned with the line joining the source and detector, which points to source. The ADM mass of boosted star or black hole is \cite{Gourgoulhon_2011}
\begin{equation}
M_{ADM}=\frac{M_0}{\sqrt{1-v^2}}=\gamma M_0\approx M_0+\mathcal{O}(v^2),
\end{equation}
where $M_0$ is the ADM mass of motionless star or black hole, and $v$ is the speed of star relative to asymptotic infinite observer. When  GW source is moving with velocity $\bm{v}$, the frequency of GW that detector receives is
\begin{equation}
f=f_0\frac{\sqrt{1-v^2}}{1+\bm{v}\cdot\bm{n}}\approx f_0(1-\bm{v}\cdot\bm{n})+\mathcal{O}(v^2),
\end{equation}
where $f_0$ is intrinsic frequency. Here we have assumed that $|\bm{v}|\ll 1$.  And the corresponding Doppler redshift is
\begin{equation}
z=\frac{f_0-f}{f}=\frac{1+\bm{v}\cdot\bm{n}}{\sqrt{1-v^2}}-1\approx \bm{v}\cdot\bm{n}.
\end{equation}

The waveform of GW produced by binary compact objects in circular orbit, in the Newtonian Limit, reads $h_{+,\times}=hf_{+,\times}(\iota)e^{i\Phi_{+,\times}(t)}$, where
\begin{equation}
h=\frac{4G^{5/3}\pi^{2/3}\mathcal{M}_0^{5/3}f_0^{2/3}}{D_0},
\label{eq:4}
\end{equation}
is the amplitude of GW, $f_{+,\times}(\iota)$ is a function with respect to the angle between orbital angular momentum of GW source and the line of sight, and 
\begin{equation}
\Phi_{+,\times}(t)=\phi_{c,+,\times}-(\frac{t_c-t}{5G\mathcal{M}_0})^{5/8},\label{eq:5}
\end{equation}
is phase. In equation~(\ref{eq:4}-\ref{eq:5}), $\mathcal{M}_0$ is chirp mass, which is equal to $\frac{(m_1m_2)^{3/5}}{(m_1+m_2)^{1/5}}$ and $m_{1},m_{2}$ are the mass of binary compact objects respectively, $f_0$ is frequency of GW, $D_0$ is the distance between detector and GW source and $\phi_c,t_c$ are two parameters. Once cosmological redshift is considered, the frequency and amplitude of GW will become $f_z=\frac{f_0}{1+z_c}$ and $h=\frac{4\pi^{2/3}\mathcal{M}_{z}^{5/3}f_z^{2/3}}{D_L}$ respectively \cite{Krolak_1987}, where $\mathcal{M}_z=\mathcal{M}_0(1+z_c)$, which means $m_{iz}=m_{i0}(1+z_c)$, $D_L=D_0(1+z_c)$, $D_L$ is luminosity distance and $z_c$ stands for cosmological redshift. $\mathcal{M}_0,f_0,D_0$ in equation~(\ref{eq:4}) are replaced by $\mathcal{M}_z,f_z,D_z$. Under this replacement, the phase of GW becomes $\Phi_{+,\times}(t)=\phi_{c,+,\times}-(\frac{(1+z)t_c-t}{5G\mathcal{M}_z})^{5/8}$, which ensures the phase of GW invariant.

When the Doppler effect is concerned, $h=\frac{4\pi^{2/3}\mathcal{M}_z^{5/3}f_z^{2/3}}{D_z}$, where $\mathcal{M}_z=\mathcal{M}_0(1+z_c)(1+z_d), f_z=\frac{f_0}{(1+z_c)(1+z_d)}$, $D_z=D_L(1+z_d)$. $z_d$ stands for Doppler redshift. For $\bm{v}\cdot\bm{n} \in[0,\pm 5000km/s]$, $z_d\in [0,\pm 0.0167]$. If the GW source is in a deep potential well, such as in the vicinity of supermassive black hole or intermediate mass black hole, then there are two kinds of redshift must be concerned. First one is the Doppler redshift. The speed of GW source revolving around the central HB, in Newtonian limit, is $v=\left(\frac{G M_{BH}}{r}\right)^{1/2}$, where $M_{BH}$ is the mass of central black hole, $r$ is the GW source orbit radius and the minimum orbit radius we assigned is $6GM_{HB}$, which is innermost stable circular orbit. So $z_d=\bm{v}\cdot\bm{n}\in[0,\pm\sqrt{1/6}]$. And the second one is gravitational redshift, which is $z_g=\frac{G M_{BH}}{r}\in[0,1/6]$. These two kinds of redshift are bigger than the redshift caused by radiation reaction.

So far, three kinds of redshift have been discussed, namely cosmological redshift $z_c$, Doppler redshift $z_d$ and gravitational redshift $z_g$. Assuming these three kinds redshift are existent at the same time, then $h$ is still expressed as $\frac{4\pi^{2/3}\mathcal{M}_z^{5/3}f_z^{2/3}}{D_z}$, and $\mathcal{M}_z=\mathcal{M}_0(1+z_c)(1+z_d)(1+z_g)$, $f_z=\frac{f_0}{(1+z_c)(1+z_d)(1+z_g)}$, $D_z=D_L(1+z_d)(1+z_g)$. It is not hard to find $D_z$ is $(1+z_d)(1+z_g)$ times as far as the true value $D_L$. Theoretically speaking, the intrinsic mass of GW source is $\frac{m_z}{(1+z_c)(1+z_d)(1+z_g)}$. But $z_c$ of GW source can not be obtained from GW signals, and it is obtained by the relation between luminosity distance and cosmological redshift in cosmology, which is $D_L=\frac{(1+z_c)}{H_0}\int_0^{z_c}\frac{dx}{E(x)}$. If $z_d$ and $z_g$ are not zero, but we ignore them and think all redshift is caused by cosmic expansion, then the redshift ($z$ in the r.h.s. of equation~(\ref{eq:6})) gotten by the relation between luminosity distance and cosmological redshift is not equal to the true cosmological redshift $z_c$. When Doppler and gravitational redshift are concerned, then
\begin{equation}
D_z=\frac{(1+z_c)(1+z_i)}{H_0}\int_0^{z_c}\frac{dx}{E(x)}=\frac{(1+z)}{H_0}\int_0^{z}\frac{dx}{E(x)},
\label{eq:6}
\end{equation}
where $1+z_i=(1+z_d)(1+z_g)$, $E(x)=\sqrt{\Omega_{m0}(1+x)^3+\Omega_\Lambda}$, $\Omega_{m0}$ is the present energy density of dust matter in the units of critical density and $\Omega_{\Lambda}$ is present energy density of dark energy. One can get redshift $z$ by solving above equation and, obversely, it is not equal to $(1+z_c)(1+z_d)(1+z_g)-1$. The mass extracted from GW is $\frac{m_z}{1+z}$, which is different from the intrinsic mass $\frac{m_z}{(1+z_c)(1+z_d)(1+z_g)}$.

\section{Damping of GW in viscous fluid}
\label{sec:3}

Wave may be dissipated and dispersed by medium, and GW makes no exception \cite{Baym_2017}. Not only viscous medium, but dust may damp the intensity of GW by scattering \cite{Sv_tek_2009}. Next we discuss the damping of GW in viscous medium.

The energy momentum tensor of adiabatic nonideal fluid is \cite{Goswami_2017}
\begin{equation}
T_{\mu\nu}=(\rho+p)u_\mu u_\nu+p g_{\mu\nu}-2\eta\sigma_{\mu\nu}-\xi\Theta\Delta_{\mu\nu}
\end{equation}
where $\Delta_{\mu\nu}=g_{\mu\nu}+u_\mu u_\nu$,$\Theta=\nabla_\alpha u^\alpha,\sigma_{\mu\nu}=\nabla_{(\mu}u_{\nu)}+a_{(\mu}u_{\nu)}-\frac{1}{3}\Theta \Delta_{\mu\nu}$, $u^\alpha,a^\alpha$ are 4-velocity and 4-acceleration respectively, $\eta$ is shear viscosity and $\xi$ is bulk viscosity. The line element that contains tensor perturbation is
\begin{equation}
ds^2=-dt^2+(\delta_{ij}+h_{ij})dx^idx^j
\end{equation}
where $h_{ij}$ are transverse and traceless, namely $\partial^jh_{ij}=0,h^i_i=0$. Here we have supposed that the background is Minkowski and wave propagates in z-direction. According to perturbed Einstein equation $\delta G_{\alpha\beta}=8\pi G\delta T_{\alpha\beta}$, the wave equation in viscous medium became \cite{Baym_2017,Goswami_2017,Kocsis_2008}
\begin{equation}
\frac{\partial^2}{\partial t^2}h_{ij}+2\beta\frac{\partial}{\partial t}h_{ij}-\frac{\partial^2}{\partial z^2} h_{ij}=0\label{eq:9}
\end{equation}
where $\beta=\frac{8\pi G\eta}{c^2}$(in SI units). Because of the existence of $\beta$, GW can be damped by viscous medium. However in general, $\beta$ is very small, so GW can pass through most matter without any observable damping.

Modern cosmology predicts that about $25\%$ matter is in the from of dark matter, which plays crucial role in cosmic evolution and the structure formation of galaxy  \cite{Salucci_2019,Borriello_2001,Gentile_2004}.  And it's generally considered that dark matter is cold, namely it's collisionless. The effect of cold dark matter on GW is too small to be detected \cite{Flauger_2018}. Self interacting dark matter (SIDM) model is used to explain the small scale observation that is inconsistent with the cold dark matter model predicted \cite{Tulin_2018}. SIDM may be the origin of the viscosity of cosmic medium. From the perspective of statistical mechanics, the viscosity of fluid comes from the distribution function of particles deviating from equilibrium distribution. In one case, GW disturbs dark matter and results in the distribution deviating from equilibrium distribution, but in this case, the damping of GW can be neglected \cite{Baym_2017}. In another case, the dark matter itself is in non-equilibrium state. And according to statistical mechanics, the shear viscosity coefficient, in the non-relativistic limit, is $\eta=\frac{1.18m<v>^2}{3<\sigma v>}\approx \frac{1.18m<v>}{3<\sigma>}$ \cite{Atreya_2019}, where $m$ is the mass of dark matter particle and $\sigma$ is scattering section of dark matter. On galaxy scale, $<\frac{\sigma}{m}>$ is about $2cm^2/g$ \cite{PhysRevLett.116.041302} and $<v>$, in general, is about $10^{-2}c$, then $\eta\approx 10^7 Pa\cdot s$. And the relaxation time $\tau\approx\frac{1}{n<\sigma v>}\approx\frac{m}{\rho <\sigma><v>}\approx 10^8 years $, which means the dark matter would be in non-equilibrium state for a long time. In general, we assume the relation between the shear viscosity and the density of dark matter is $\eta=k\rho^\lambda=k\rho_0^\lambda(\frac{\rho}{\rho_0})^\lambda$ \cite{Velten_2014,Brevik_2019}. When GW source is surrounded by high density dark matter, the damping of GW may be observable. Dark matter Mini-halo or "mini-spike" provides a ideal approach to probe the properties of dark matter \cite{PhysRevD.72.103517}.

If a large compact object, such as massive black hole, exists in dark matter halo, then it will change surrounding dark matter distribution. A special case is that intermediate mass black hole (IMBH)($10^2M_{\sun} -10^6M_{\sun}$) exists in mini-halo and "mini-spike" is produced around the black hole \cite{Eda_2013}. The density profile of mini-halo, before the spike formed, is \cite{Mondal:2020iet}
\begin{equation}
\rho_{NFW}(r)=\frac{\rho_s}{(r/r_s)(1+r/r_s)^2},
\end{equation}
where $\rho_s,r_s$ are two parameters and more  information about these two parameter please refer \cite{PhysRevD.72.103517}. The density profile of the "mini-spike" around the central IMBH is \cite{PhysRevD.91.044045}
\begin{equation}
\rho(r)=\left\{ \begin{array}{cc}
\rho_{sp}(\frac{r_{sp}}{r})^{7/3}, &r_{min}\leq r\leq r_{sp}; \\
\rho_{NFW}(r), & r>r_{sp},
\end{array}\right.
\end{equation}
where $r_{min}=6GM_{BH}$ is innermost stable circular orbit and $M_{BH}$ is the mass of central BH. The parameters of the model we used are: $M_{BH}=10^3M_{\sun}$, $\rho_{sp}=226M_{\sun}/pc^3,r_{sp}=0.54pc,r_s=23.1pc$ \cite{Yue_2018}. If a compact subject revolves around the central BH with "mini-spike", this binary systerm will be a good GW source. The GW coming form this source is different from the source that the central BH is without "mini-spike", so this provides a new way to probe dark matter \cite{Yue_2019,Macedo_2013}.

\begin{figure}
\includegraphics[width=\columnwidth]{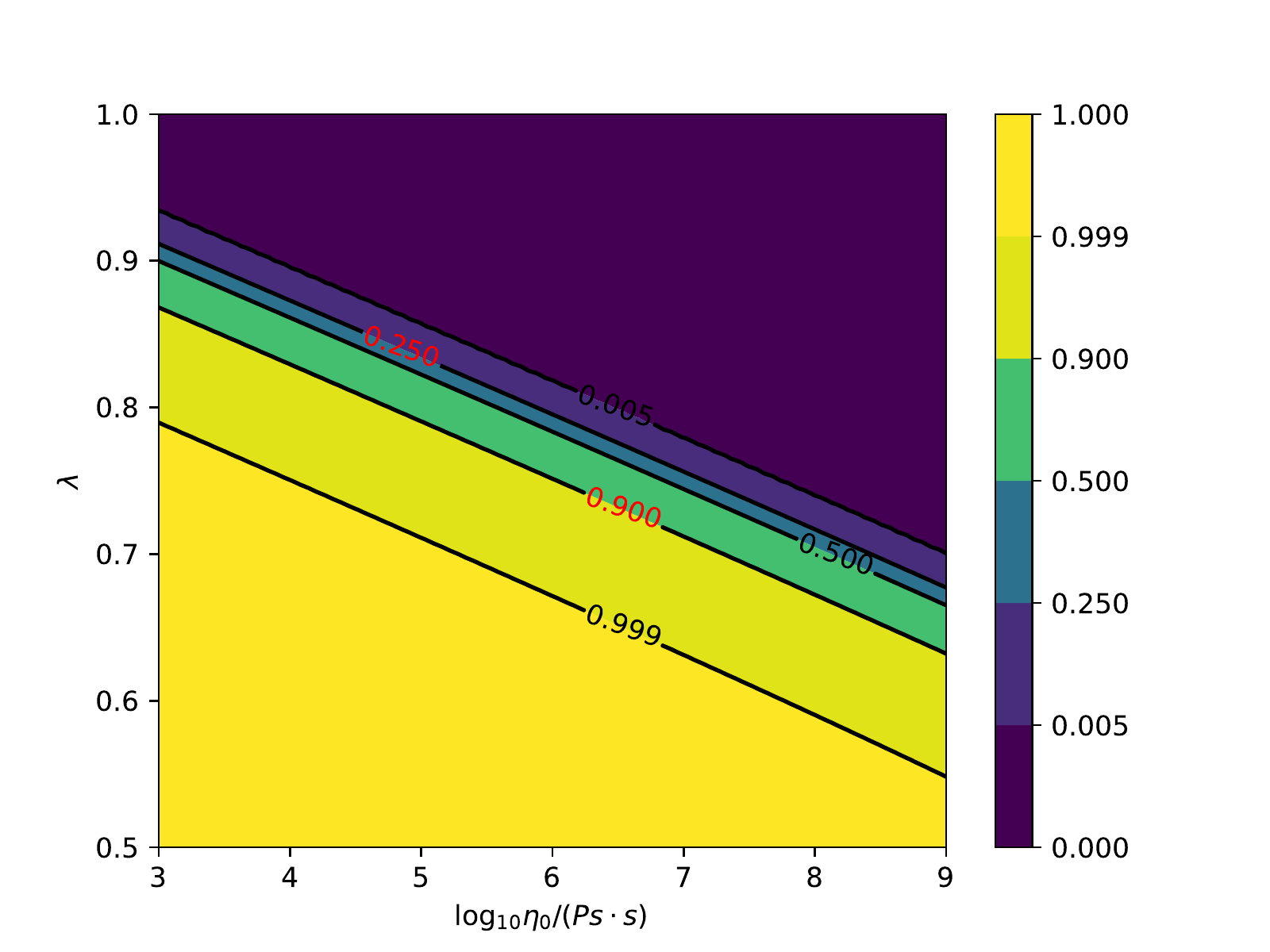} 
\includegraphics[width=\columnwidth]{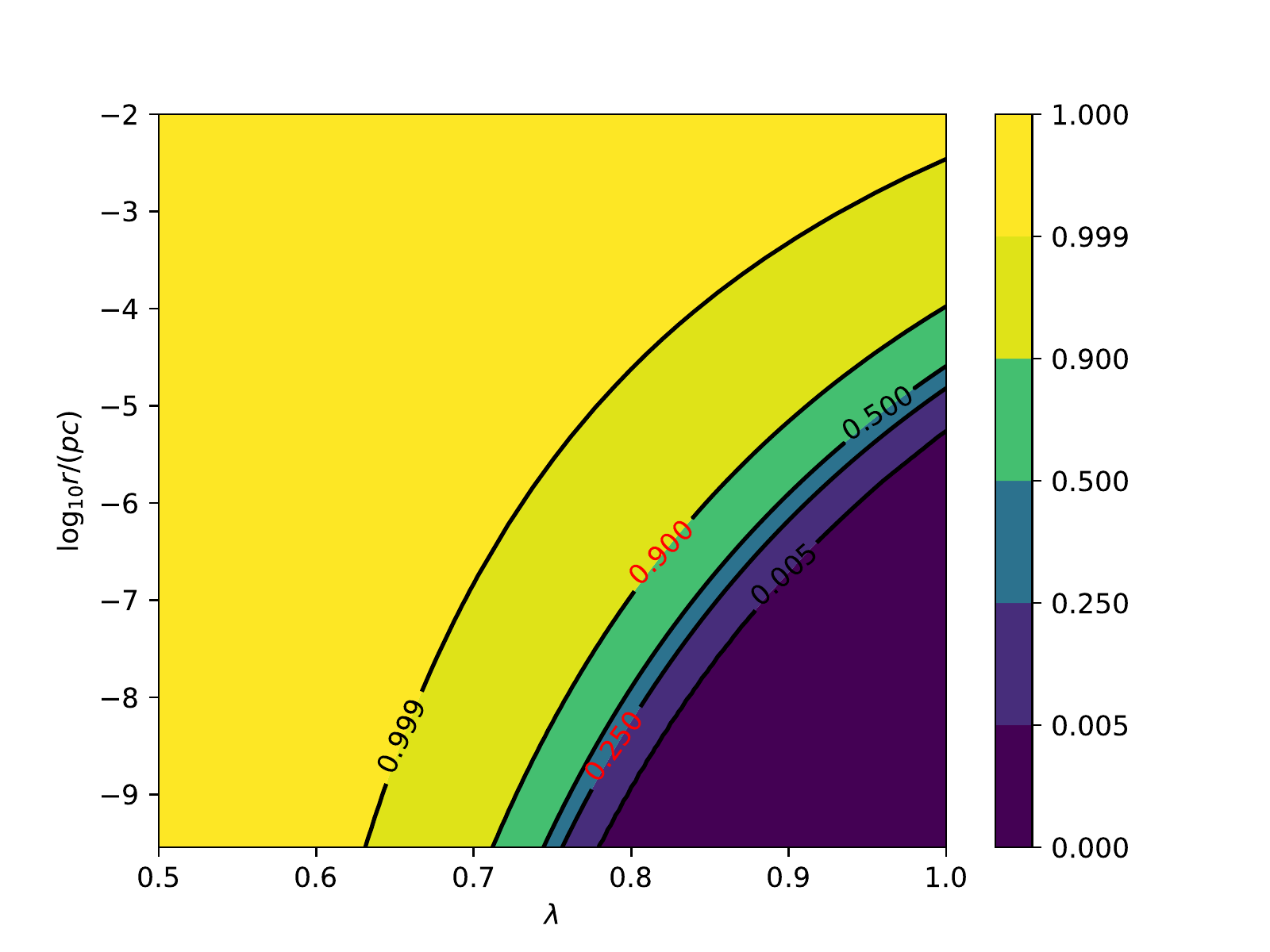} 
\caption{This two panels are the contour plot of $\frac{1}{1+z_v}$. In the upper panel, the horizon axis is $\eta_0$ and the vertical axis is $\lambda$. The GW source robit radius is $r=r_{min}$. In lower panel, the horizon axis is robit radius $\lambda$ and the vertical axis is $r$. $\eta_0$ in this case is $10^7 Pa\cdot s$.
\label{fig1}}
\end{figure}

We assume the GW source revolves around the IMBH with "mini-spike", so we can estimate the radial damping of GW's amplitude. The approximate solution of equation~(\ref{eq:9}) is \cite{Goswami_2017}
\begin{equation}
h_{ij}\approx h e^{-\beta t-i\sqrt{k^2-\beta^2}t+i kz}\approx h e^{-\int\beta dx-ikt+i kz}.
\end{equation}
The term $e^{-\int\beta dx}$  stands for the damping and

\begin{equation}
\int\beta dx\approx\frac{8\pi G}{c^3} k\rho_0^{\lambda}(\frac{\rho_{sp}}{\rho_0})^\lambda [\int_{r}^{r_{sp}} (\frac{r_{sp}}{x})^{\frac{7\lambda}{3}}dx
\end{equation}
where $r$ is the GW source orbit radius and the following calculation shows $\int_{r_{sp}}^\infty\beta dx\approx 0$. Here, we define $z_v=e^{\int\beta dx}-1$, then $e^{-\int\beta dx}=\frac{1}{1+z_v}$. One can name $z_v$ viscosity redshift, and this redshift has no effect on the GW frequency. There are three parameters $\rho_0,\eta_0=k\rho_0^\lambda,\lambda$ in above equation, and the effect of these three parameters on $z_v$ have been plotted in Fig.~\ref{fig1}. Since the information about $<\sigma/m>$ on mini-halo scale is insufficient and the interaction between dark matter particles is not just elastic collision in high density region \cite{PhysRevD.72.103517,Atreya_2019,Goswami_2017}, we have assumed $\eta=\eta_0(\frac{\rho}{\rho_0})^\lambda$. For $\rho=\rho_0\approx 10^{-22}kg/m^3$, which is approximate to the average density of dark matter on galaxy scale, $\eta=\eta_0\approx 10^{7}Pa\cdot s$. So we set the range of the parameters as: $\rho_0=10^{-22}kg/m^3,\eta_0=k\rho_0^\lambda=[10^3,10^9]Pa\cdot s,\lambda=[0.4,1],r=[r_{min},0.01]pc$. 

It is not hard to find the damping is very intense if the parameters are at certain range.  According to Fig.~\ref{fig1}, the damping is almost negligible when $\lambda <0.6$ or $r>0.01pc$. And $\frac{1}{1+z_v}$ is zero when the parameters are at certain range, which means that $z_v$ is infinite and the GW is totally damped by dark matter. 
When viscosity redshift $z_v$ is concerned, the amplitude of GW becomes $\frac{4\pi^{2/3}\mathcal{M}_z^{5/3}f_z^{2/3}}{D_z}$, where $\mathcal{M}_z=\mathcal{M}_0(1+z_c)(1+z_d)(1+z_g)$, $f_z=\frac{f_0}{(1+z_c)(1+z_d)(1+z_g)}$, $D_z=D_L(1+z_d)(1+z_g)(1+z_v)$. The luminosity distance increases by $(1+z_v)$ times. Again, one can get redshift $z$ by taking advantage of equation~(\ref{eq:6}), and in this case, $(1+z_i)$ in equation~(\ref{eq:6}) is $(1+z_d)(1+z_g)(1+z_v)$. The mass extracted from GW is $\frac{\mathcal{M}_z}{1+z}$, which is different from the intrinsic mass $\frac{m_z}{(1+z_c)(1+z_d)(1+z_g)}$.
\begin{figure}
\includegraphics[width=\columnwidth]{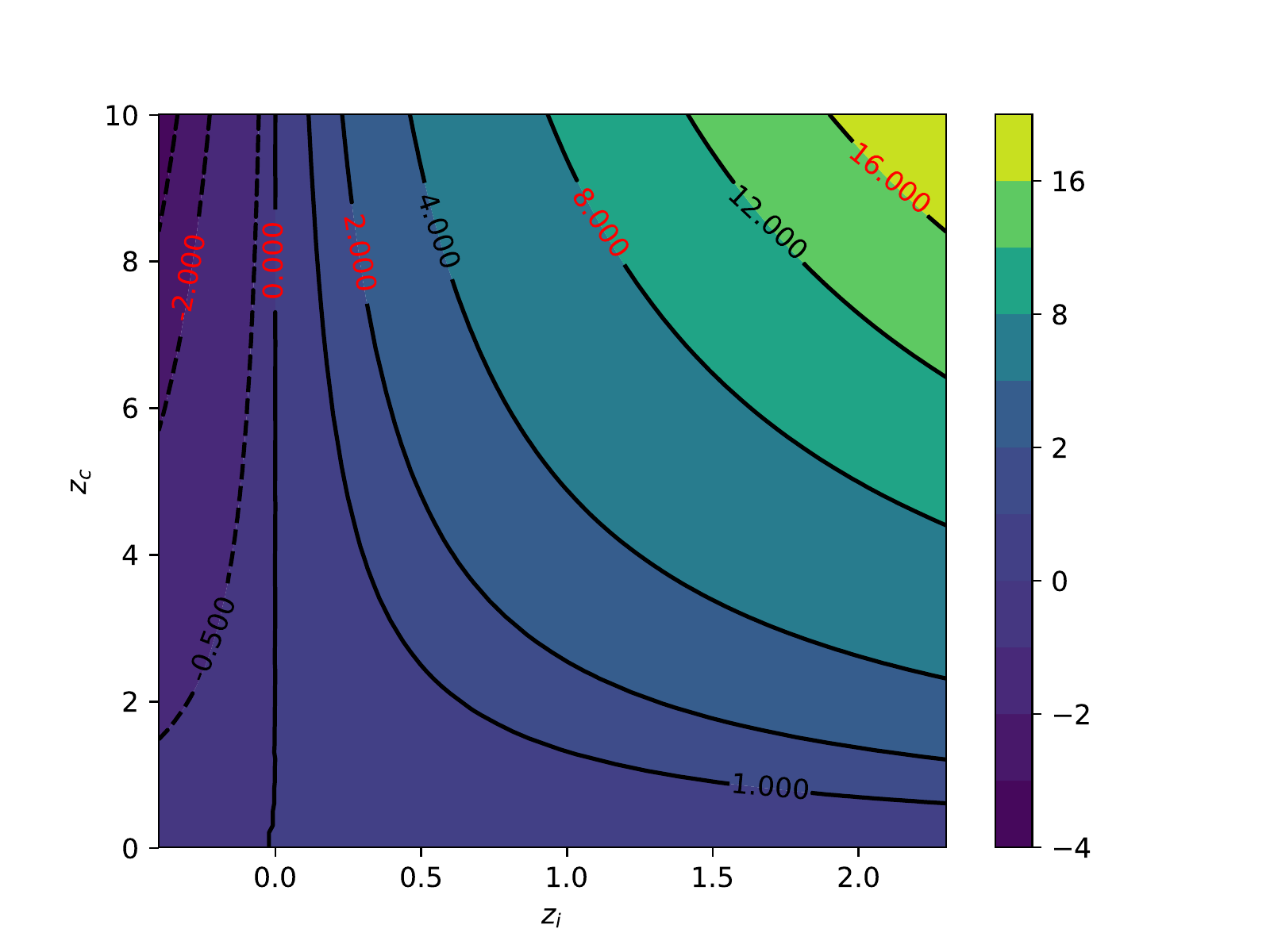}
\includegraphics[width=\columnwidth]{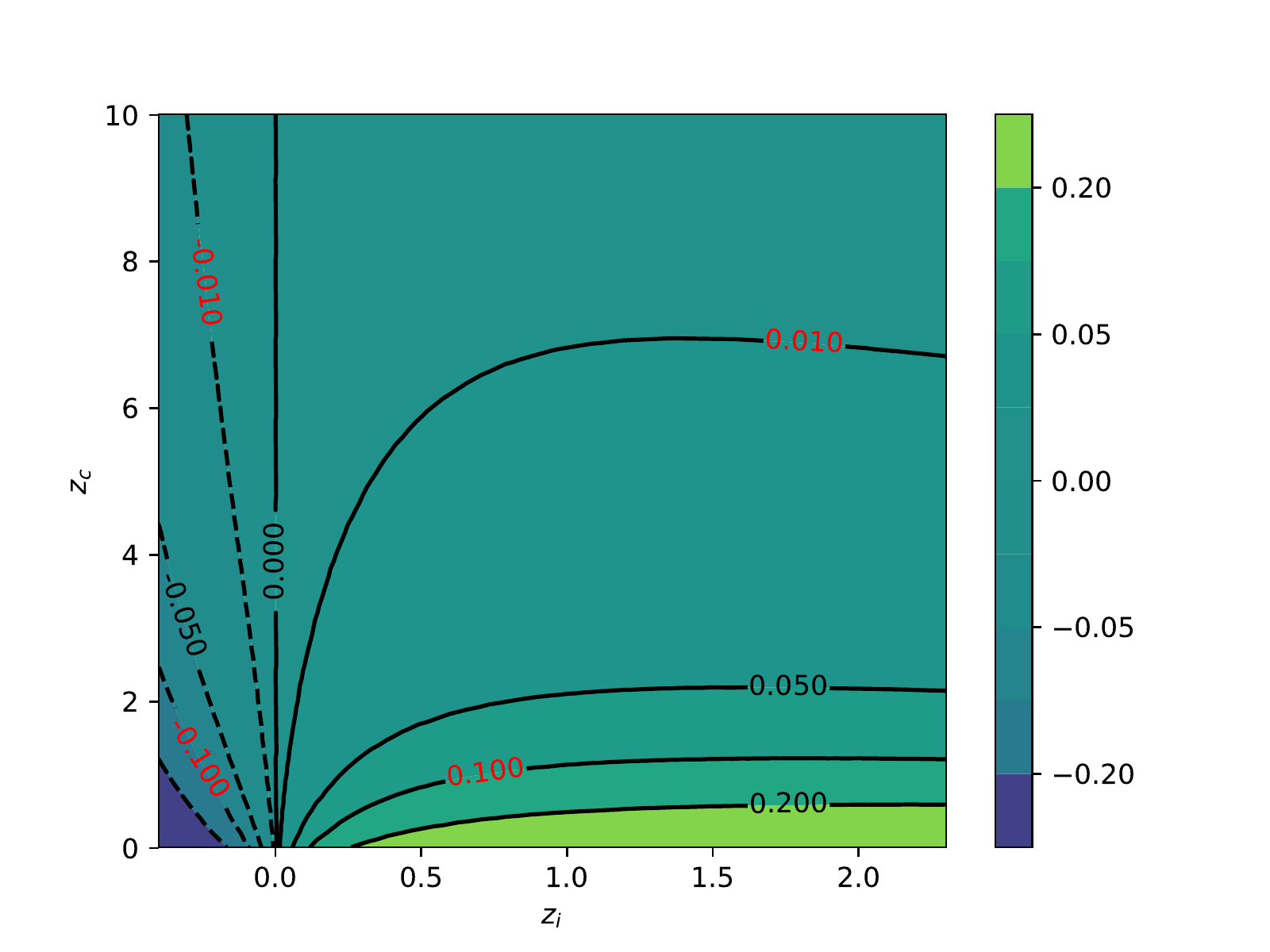}  
\caption{The upper panel is the contour plot of $z-z_c$. The lower panel is the contour plot of $\frac{1}{1+z}-\frac{1}{(1+z_c)(1+z_d)(1+z_g)}$, which is in units of $m_z$. \label{fig2}}
\end{figure}

According to the calculation of previous section, the biggest Doppler redshift is up to $\pm 0.4$, which is almost the biggest cosmological redshift of known GW source \cite{GWE}. And the biggest gravitational redshift is up to $0.167$, which is bigger than the cosmological redshift of some GW source. Here we assume the biggest $z_v$ is $1$, which means half of GW amplitude is damped. So the range of $z_i$ , which is equal to $(1+z_d)(1+z_g)(1+z_v)-1$, is $[-0.4,2.3]$. The contour plots of $z-z_c$ and $\frac{m_z}{1+z}-\frac{m_z}{(1+z_c)(1+z_d)(1+z_g)}$ with respect to $z_c$ and $z_i$ have been shown in Fig.~\ref{fig2}. It is not hard to find the effect of $z_i$ on $z$ is very intense when $z_c$ is high and the difference between $z$ and $z_c$ may be bigger than $z_c$ itself, which means in extreme situation the whole procedure of redshift estimation from GW signal may become meaningless. And the effect of $z_i$ on mass $m$ is intense if $z_c$ is low and $\frac{m_z}{1+z}$ is less than $\frac{m_z}{(1+z_c)(1+z_d)(1+z_g)}$ for $z_i<0$.

The viscosity of cosmic medium, if it exists, is more important because all GWs would be affected by it. Beside that the viscosity would play very important roles in cosmic evolution \cite{Schatz:2016gsy,Velten_2014}, it even may play the role of dark energy, which causes the cosmic acceleration \cite{Atreya_2019}. In above discussion, we regard $z_v$ as a constant, but if the damping of GW is caused by cosmic medium, then $z_v\propto D_L$, so $z_v$ is function of $z_c$. Current ground-based GW detector can dectect the GW source at $z_c=1$, but third-generation detector, such as Einstein Telescope or Cosmic Explorer, may detect GW sources at $z_c=10$ \cite{Vitale_2019}, or even at $z_c=100$ \cite{Hall_2019,Baibhav_2019}. So whether the viscosity of cosmic medium is existent will be a crucial problem. If it do exists, then the luminosity distance and mass of GW source extracted from GW should be corrected accordingly, especially when the GW source is at high redshift.

Merger rate and detection rate of compact objects may provide a way to detect the the viscosity of cosmic medium. Merger rate and detection rate of compact objects are very important conceptions in cosmology \cite{Rodriguez_2018}, which carry the information about the evolution of compact objects. The relation between detection rate and merger rate is \cite{Dominik_2015}
\begin{equation}
R_{det}=\iiint_0^{\infty} p_{det}\mathcal{R}(z_m)\frac{dV_c}{dt_m}\frac{dt_m}{dt_{det}}dz_{m}dm_1dm_2
\end{equation}
where $R_{det}$ is detection rate, $\mathcal{R}(z_m)$ is merger rate and  $\mathcal{R}(z_m)=\frac{dN}{dm_1dm_2dV_cdt_m}$ , $p_{det}=p_{det}(z_m,m_1,m_2)$ is detectable probability. In local universe, merger rate of BBH $\mathcal{R}=53.2^{+58.5}_{-28.8}Gpc^{-3}yr^{-1}$ \cite{Abbott_2019} and BNS $\mathcal{R}=1540^{+3200}_{-1220}Gpc^{-3}yr^{-1}$ \cite{Howell_2019}. Merger rate provides a way to probe the information of the environment in which compact objects formed and the BBH formation channel \cite{Fishbach_2018,Zevin_2017,Belczynski_2002}. Moreover merger rate can be used to constrain astrophysical model \cite{Vitale_2019}. Primordial black holes (PBHs) are candidate of dark matter, so investigating the merger rate of PBHs is an important way to study dark matter \cite{Chen_2018,Gow_2020,Raidal_2019}. In a word, precise evolution of merger rate is crucial cosmology and astrophysics. As the increasing of the detector sensitivity, more and more GW events will be detected, which will  further contribute to the estimation of high-precision merger rate \cite{Abbott_2019,Mapelli_2018}.

If the viscosity of cosmic medium is existent, even if it is small, the damping rate of GW is considerable because of the long propagation distance, and the merger rate extracted from GW events must differ from the rate obtained by other approaches, such as numerical simulation \cite{Micic_2007,Barrett_2018} or other observation \cite{Evibh,10.1093/mnras/stu786}. So in order to get high precision merger rate of the compact objects, whether the viscosity of cosmic medium is existent will be a crucial factor. Conversely, the merger rate can be used to constraint the viscosity of cosmic medium.

\section{Conclusion}

In this article, we introduced three kinds of redshift, which are Doppler redshift $z_d$, gravitational redshift $z_g$ and viscosity redshift $z_v$ and discussed how they "contaminate" the GW signal. These three kinds of redshift are not concerned in the standard procedure of GW data processing. So the parameters estimated by GW, such as the luminosity distance, redshift and intrinsic mass, may intensely deviate from their intrinsic values. So it is important to confirm whether these redshift are existent. In turn, if the information of GW source is known in advance, then one can probe the properties of cosmic medium (or dark matter) by GW signal.

\section{Acknowledgments}
This article is supported by the National Natural Science Foundation of China (Grant No.11675032,12075042).

\section*{Data Availability}
The data underlying this article will be shared on reasonable request to the corresponding author.

\bibliographystyle{mnras}
\bibliography{referencelist}

\bsp	

\label{lastpage}
\end{document}